\documentclass[showpacs,prb,aps,twocolumn]{revtex4}
\usepackage{graphicx}
\usepackage{color}
\usepackage{dcolumn}
\usepackage{bm}
\usepackage{amsmath}
\usepackage{amssymb}
\def\n{\mathbf{n}}

\def\gapx{\lower 2pt \hbox{$\buildrel>\over{\scriptstyle{\sim}}$\ }}
\def\lapx{\lower 2pt \hbox{$\buildrel<\over{\scriptstyle{\sim}}$\ }}
\def\j1j2{$J_1$$-$$J_2$}
\begin{document}
\title{Ground-state phase diagram of the quantum $J_1-J_2$  model on the square lattice}

\author{Fabio Mezzacapo}

\affiliation{$$Max-Planck-Institut f\"ur Quantenoptik, Hans-Kopfermann-Strasse 1, D-85748, Garching, Germany}
\date{\today}

\begin{abstract}
We study the ground-state phase diagram of the quantum $J_1-J_2$ model on the square lattice by means of an entangled-plaquette variational ansatz. In the range  $0\le {J_2}/{J_1} \le 1$, we find  classical magnetic order of  N\'eel and collinear type, for ${J_2}/{J_1}\lesssim 0.5$,  and   $J_2/J_1 \gtrsim 0.6$ respectively. For intermediate values of $J_2/J_1$ the ground state is a spin liquid (i.e., paramagnetic with no valence bond crystalline order). Our estimates of the entanglement entropy show that such a spin liquid is topological. 
 \end{abstract}


\maketitle
\section{Introduction}
The spin-1/2 antiferromagnetic (AF) Heisenberg Hamiltonian with additional AF next-nearest neighbor coupling is one of the most studied models, known as the $J_1-J_2$, in theoretical physics:
\begin{equation}
H=J_1 \sum_{\langle i,j \rangle}\mathbf{S}_i \cdot \mathbf{S}_j +J_2\sum_{\langle\langle i,j \rangle\rangle}\mathbf{S}_i \cdot \mathbf{S}_j.
\label{ham}
\end{equation}
Here ${\bf S}_i$ is a spin-1/2 operator associated to the $i$th lattice site and the first (second) summation runs over nearest (next-nearest) neighbor sites.\cite{note}

Model (\ref{ham}) is of relevance for experimentally accessible compounds,\cite{exp1, exp2} and constitutes a canonical example of spin system featuring frustration induced by competing AF interactions. 
It is known that, on the square lattice, the ground state (GS) of (\ref{ham})  displays AF long-range order of N\'eel [ordering wave vector $\mathbf{q}=(\pi, \pi)$]  type at $J_2=0$ (i.e., in the unfrustrated case);\cite{Sand} the N\'eel order remains for sufficiently  small values of $J_2$ while at large $J_2$ the AF order is collinear [$\mathbf{q}=(\pi, 0)$ or $(0,\pi)$].\cite{IM} To gain insight into the nature of the GS at intermediate values of $J_2$, several numerical techniques have been applied. The exact diagonalization (ED) of the Hamiltonian (\ref{ham}) is possible only for small lattice size (up to $N\sim 40$ spins), therefore, although this approach may give fundamental suggestions, a reliable extrapolation of the relevant physical observables to the thermodynamic limit is basically not achievable. On the other hand, quantum Monte Carlo (QMC) schemes based on the imaginary time evolution of a trial wave function (WF) are essentially exact, even for very large lattices, at $J_2=0$, being affected by the sign problem at finite $J_2$ (i.e., in presence of frustration). Variational Monte Carlo (VMC) calculations overcome the major limitations of ED and QMC being not restricted to small lattice size and, by definition, free of any sign instability. However, their accuracy  depends on the guess for the GS WF. Hence, understanding and characterizing the GS properties of the $J_1-J_2$ model in the maximally frustrated regime (i.e., $J_2\sim0.5$ for a square geometry) is as difficult as  it is interesting. 

ED findings are consistent with a paramagnetic valence bond crystalline (VBC) phase intervening between the N\'eel and collinear one.\cite{edsq} This conclusion also arises (albeit with possibly different VBC patterns) from recent  projected-entangled pair states,\cite{IM} series expansion and spin-wave theory,\cite{Oitmaa} coupled-cluster,\cite{CCM} hierarchical mean field,\cite{Ortiz} and renormalized tensor network\cite{ren} investigations,  whose main objective has been that of elucidating the type of phase transition occurring at the boundary between the N\'eel and paramagnetic VBC state. 

Recent VMC calculations pointed out the existence of a spin liquid (i.e., paramagnetic with no VBC order) GS for model (\ref{ham}) on a honeycomb lattice,\cite{clark, epshoney} revitalizing the long-lasting debate concerning the possibility of observing a similar phase on the square lattice.\cite{Sor, dmrg, peps}  

In this paper we investigate the GS phase diagram of (\ref{ham}) using a VMC approach based on the variational family of entangled-plaquette states (EPS).\cite{eps} The EPS WF is a general ansatz which has been applied to a variety of unfrustrated (bosonic) or frustrated (fermionic) lattice problems yielding results of accuracy at least comparable to (or better than) that obtainable with different numerical schemes or alternative variational WF's.\cite{epshoney, eps, epscps} In particular it has been employed with success to determine the GS phase diagram of  (\ref{ham}) on the honeycomb lattice,\cite{epshoney} giving predictions quantitatively and qualitatively more accurate than those affordable using different ansatzs.\cite{clark} The main quality of the EPS WF is its systematic improvability which allows one to compute increasingly accurate estimates of physical observables by sequentially increasing the plaquette size (i.e., the number of spins comprised in a single plaquette) $l$.

 We estimate GS energies and relevant order parameters, i.e., square sublattice magnetization (SSM) and various VBC structure factors, in the range $0\le J_2 \le 1$. For each system size, up to $N=L\times L=256$ spins with periodic boundary conditions (PBC),  the EPS WF is independently optimized, starting with $N$ square plaquettes of four sites. The size of our plaquettes is then consistently and systematically increased up to sixteen spins by adding $2l^{1/2}+1$  sites each time  that $l$ is changed. We extrapolate the estimates obtained on finite lattices of different sizes, for a given plaquette size,  to the thermodinamic limit and, finally,  asses the dependence of our results on $l$. 
 
Our main findings are the following: {\it a}) the system is N\'eel ordered up to $J_2 \sim 0.45$, while for $J_2\gtrsim 0.6$ the magnetic order is collinear; {\it b}) for intermediate values of $J_2$ all the order parameters considered here vanish in the thermodynamic limit signaling the emergence of a spin liquid; {\it c}) by estimating the entanglement entropy\cite{ign, melko1, melko2} we characterize such a spin liquid as topological.

\section{The entangled-plaquette variational wave function}

In this section we briefly recall the basic aspects of the EPS WF, referring the reader to Refs. \onlinecite{eps}, and \onlinecite{epscps} for further details. 

A generic WF for a spin-1/2 lattice system, e.g.,  the one described in model (\ref{ham}), can be expressed as  
\begin{equation}
|\Psi\rangle=\sum_{\mathbf{n}}W(\mathbf{n})|\mathbf{n} \rangle
\end{equation}
where  $|\mathbf{n} \rangle = |n_1,n_2, \ldots , n_N\rangle$,  $n_i $ is the eigenvalue of $
\sigma^z_i $ and $W(\mathbf{n})$  is the weight of a configuration of the system. The  EPS ansatz  is based on the following idea: cover the lattice  with $N$ plaquettes of  $l$ sites,   and identify the weight of a given global configuration with the product of $N$ variational coefficients in one-to-one correspondence with the particular plaquette configuration. Hence, 
\begin{equation}
 \langle \n | \Psi \rangle=W(\n)=\prod_{P=1}^N C_P^{\n_{P}}
 \end{equation} 
where the configuration of the $P_{th}$ plaquette is given by the values  $\n_P=n_{1,P},n_{2,P},\ldots,n_{l,P}$ at its $l$ sites. The above choice of $W(\n)$ fully specifies the EPS WF. Such an ansatz, when overlapping (i.e., entangled) plaquettes are considered (i.e., $l >1$) naturally incorporates spin-spin correlations, a crucial ingredient to describe the physics of a quantum GS. As already mentioned, the fundamental property of this variational choice is its  systematic improvability obtainable by increasing $l$. Given its simple form, the EPS WF is efficiently optimizable on a computer; moreover, once optimized, it allows  for an evaluation of the physical observables  efficient as well. Both the WF optimization and the estimates of physical observables are achieved via the Monte Carlo method.\cite{epshoney, eps, epscps}  

\section{Results}
\begin{figure}[t]
{\includegraphics[scale=0.68]{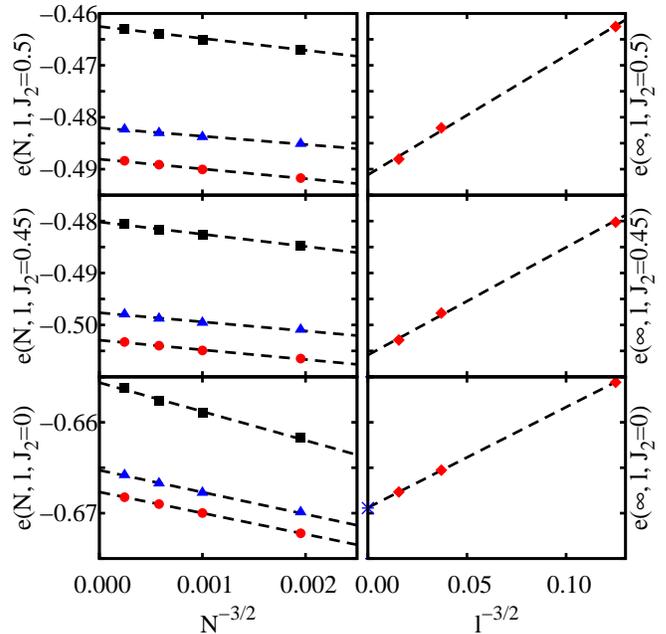}}
\caption{(color on line). Left: GS energy per site of the $J_1-J_2$ model on the square lattice with PBC, as a function of the system size. Estimates are obtained with $N$ plaquettes of $l=4$ (boxes), $l=9$ (triangles) and $l=16$ (circles) sites. Each dashed line is a fit  to numerical data with the same $l$ value. Right: GS energy per site, extrapolated to the thermodynamic limit, as a function of the plaquette size. Each dashed line is a fitting function linear  in $l^{-3/2}$ (see text). The QMC estimate in the $J_2=0$ case is also shown for comparison (star).\cite{Sand}}  
\label{fig:1}
\end{figure}

Figure \ref{fig:1} shows the GS energy per site $e=E/N$ as a function of the system size for various values of $J_2$ and different plaquette size (left part). Regardless the system size and the value of $J_2$ our estimates sensibly improve when $l$ increases. For example we obtain $e(N=100, l=4, J_2=0)=-0.65900(6)$ and  $e(N=100, l=16, J_2=0)=-0.67000(5)$ using $2\times 2$ and $4\times 4$ square plaquettes  respectively.  Estimates obtained with the same $l$ have been extrapolated to the thermodynamic limit assuming the scaling form $e(N, l, J_2)=e(\infty, l, J_2)+\alpha(l, J_2)N^{-3/2}$. It is clear that extrapolated energies depend on $l$. While in principle larger plaquette sizes have to be employed in order to observe numerical convergence of the results, available computational resources limit the largest plaquette size which can be used in practice.  However, by construction of the EPS ansatz, the estimate of any observable must approach the exact GS value in the limit of large $l$. Therefore, the point here is whether it is possible to give a reliable estimate of such a limit based on the results obtained with plaquettes of relatively small size. As pointed out in a previous work\cite{epshoney} on model (\ref{ham}) on a honeycomb lattice, this is indeed the case and estimates of the energy and all the relevant physical observables can be obtained, in the large $l$ limit, by those computed with plaquette of different size (up to $l=16$ in this study).
In the right part of Fig. \ref{fig:1} we present an example of this procedure: we expand our energy per site extrapolated to the thermodynamic limit in  powers of $l^{-1/2}$ and fit our data with the smallest number of parameters. We observe that a good description of our results is obtained employing a single power (i.e., $l^{-3/2}$). For instance, in absence of frustration, our energies $e(\infty, l, J_2=0)$ fall almost perfectly on a straight line when plotted against $l^{-3/2}$, and we find an extrapolated to the large $l$ limit energy in agreement with the QMC estimate,\cite{Sand} exact in this case.\cite{notea}
\begin{figure}[t] 
{\includegraphics[scale=0.68]{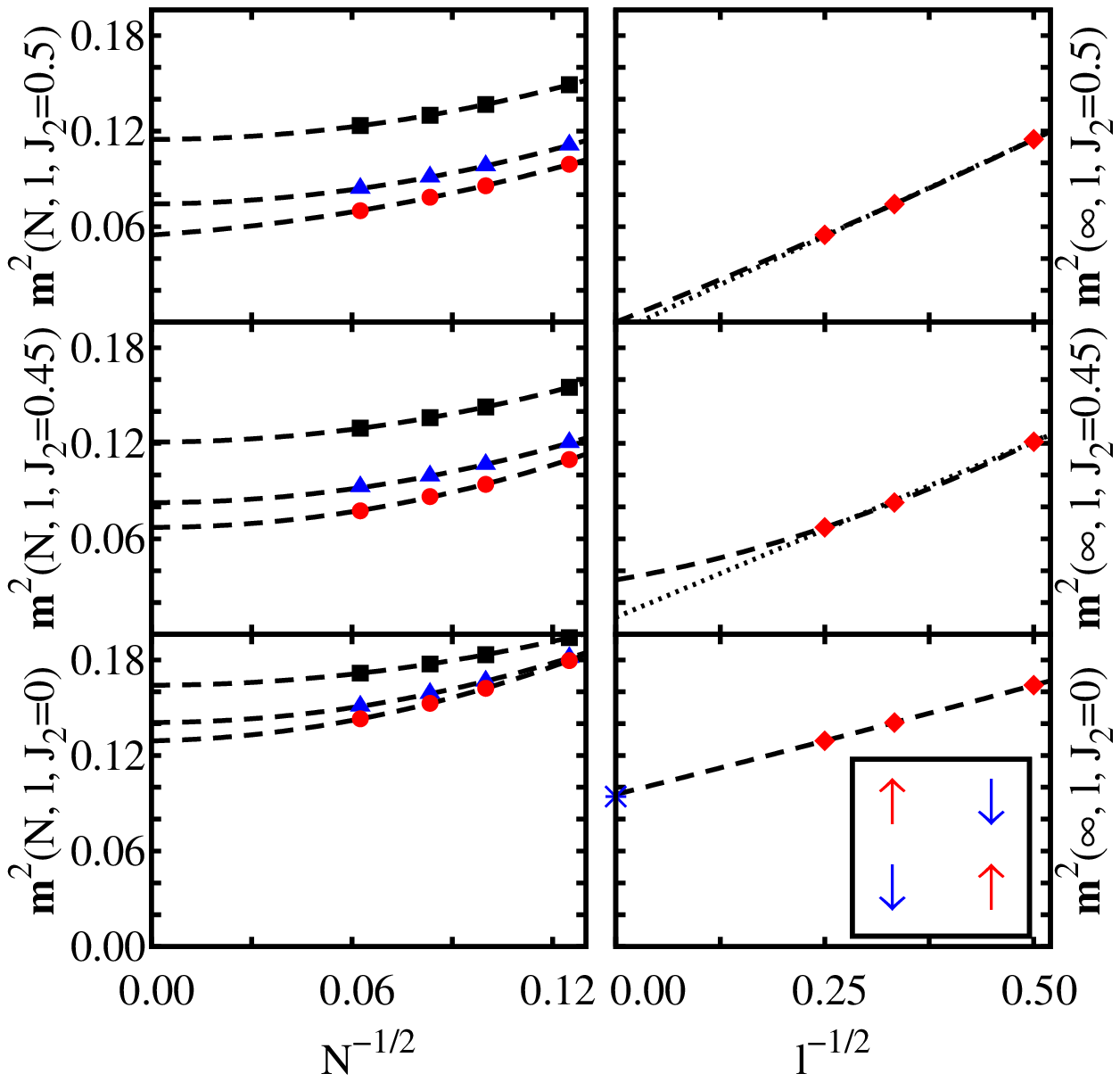}}
\caption{(color on line). Left: N\'eel  SSM of the $J_1-J_2$ model on the square lattice with PBC,  as a function of the system size. Estimates are obtained with $N$ plaquettes of $l=4$ (boxes), $l=9$ (triangles) and $l=16$ (circles) sites. Each dashed line is a fit  to numerical data with the same $l$ value (see text). Right: N\'eel  SSM, extrapolated to the thermodynamic limit, as a function of the plaquette size. Lines are functions bult to infer the $l$ dependence of our data (see text). The QMC estimate in the $J_2=0$ case is also shown for comparison (star).\cite{Sand} Inset: pictorial representation of classical N\'eel order where ``up" (``down") spins reside on sublattice $\alpha$ ($\beta$).} 
\label{fig:2}
\end{figure}

\begin{figure}[t] 
{\includegraphics[scale=0.68]{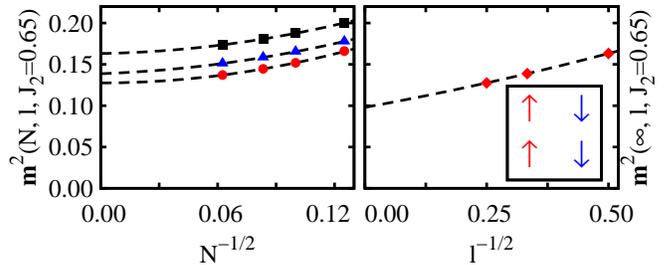}}
\caption{(color on line). Same of Fig. \ref{fig:2} where the SSM describes now magnetic order of collinear type (sublattices $\alpha$ and $\beta$ are chosen according to the inset).} 
\label{fig:3}
\end{figure}

Aimed at identifying extended regions of the phase diagram which are magnetically ordered (disordered), we compute the SSM defined as
\begin{equation}
\mathbf{m}^2(N)=\biggl \langle\frac{1}{N^2}(\sum_{i\in \alpha}\mathbf{S}_i-\sum_{j\in \beta}\mathbf{S}_j)^2\biggr \rangle
\label{eq:mag}
\end{equation}
where the two summations run over lattice sites belonging to different sublattices. In this work we consider two types of magnetic long-range order, schematically illustrated in the inset of Fig. \ref{fig:2} (N\'eel type) and Fig. \ref{fig:3} (collinear type). Specifically, if the sublattices $\alpha$ and $\beta$ are chosen  according to the inset of Fig. \ref{fig:2} (Fig. \ref{fig:3}), Eq. (\ref{eq:mag}) defines the N\'eel (collinear) SSM, or the magnetic structure factor at ordering wave vector  $\mathbf{q}=(\pi,\pi)$ ($\mathbf{q}=(\pi,0)$) 

The N\'eel SSM as a function of the system size, for different $l$ and various $J_2$ values is presented in Fig. \ref{fig:2}. For each plaquette size the extrapolation to the thermodynac limit has been performed assuming the functional form $\mathbf{m}^2(N, l, J_2)=\mathbf{m}^2(\infty, l, J_2)+\beta(l, J_2)N^{-1/2}+\delta(l, J_2)N^{-1}$. The right part of the figure shows the extrapolated values $\mathbf{m}^2(\infty, l, J_2)$ versus $l^{-1/2}$. At $J_2=0$ we find that the N\'eel SSM, in the large $l$ limit, extracted by fitting the data with a second order (in $l^{-1/2}$) polynomial, is in agreement with the QMC result.\cite{Sand} For larger $J_2$ this order parameter decreases. At $J_2=0.45$ it is still finite; this conclusion is based on the observation that both a linear (dotted line) and quadratic (dashed line) fit of the $\mathbf{m}^2(\infty, l, J_2=0.45)$ values are acceptable and yield a finite extrapolated value in the large $l$ limit.  Specifically, when two powers are considered, the visible upward bending of the data is better described, and the extrapolated value of the order parameter is slightly larger than that found with a linear fitting function. 
At $J_2=0.5$ a fit which includes only the zero-th and first order term (in $l^{-1/2})$ gives an unphysical extrapolation, while the inclusion of an extra term, proportional to $l^{-1}$, leads to a null (within the accuracy of our calculation) value of the N\'eel SSM in the large plaquette size limit. Based on this analysis it is reasonable to locate the phase boundary corresponding to  vanishing N\'eel order in the vicinity of $J_2=0.5$.

\begin{figure}[b]
{\includegraphics[scale=0.68]{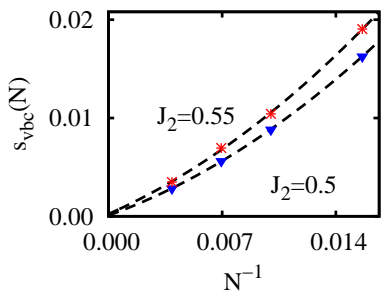}}
\caption{(color on line). Finite size scaling of the plaquette VBC order parameter for two values of $J_2$ at which the system is found in a magnetically disordered GS. Dashed lines are fits to the numerical data with the same $J_2$. This order parameter clearly vanishes, for all the $l$ values considered in this work, in the thermodynamic limit. Data shown refer to the $l=16$ case.} 
\label{fig:4}
\end{figure}

For $0.5\lesssim J_2\lesssim0.6$  no evidence of magnetic order (either of the N\'eel or collinear type) is found, while for larger $J_2$, up to $J_2=1$ (i.e., the maximum $J_2$ value studied here) the system is collinearly ordered.

The dependence of the collinear SSM on the system size is shown, for $J_2=0.65$, in Fig. \ref{fig:3} (left part). Our data extrapolated to the thermodynamic limit versus $l^{-1/2}$ are presented in the right part of the figure.  In the limit of large plaquette size the collinear order parameter stays clearly finite. Such a conclusion is consistent with both a linear and quadratic  fit of the data.

Our prediction of an extended paramagnetic region in the phase diagram of the $J_1-J_2$ model on the square lattice is in agreement with previous works carried on with different numerical approaches.\cite{IM, edsq, Oitmaa, CCM, Ortiz, ren, dmrg, peps} Most of them\cite{IM, edsq, Oitmaa, CCM, Ortiz, ren} found that VBC order, either of plaquette- or columnar-dimer type, characterizes such a magnetically disordered region.

In order to elucidate the existence of VBC order we compute, in the parameter range of interest, both the plaquette VBC and columnar-dimer structure factor defined as in Ref. \onlinecite{edsq}. Results for the plaquette VBC structure factor are presented, as a function of the system size, for two value of $J_2$ at which magnetic order is not present,  in Fig. \ref{fig:4}. The numerical data for finite system size are extrapolated to the thermodynamic limit by  means of a fitting function quadratic in the inverse of the system size.  We find that, in the limit of infinite system size, the order parameter of our interest  vanishes, within the accuracy of our calculation, regardless the plaquette size used in the EPS WF (the error on the extrapolated estimates  is of the order of $10^{-4}$). The columnar-dimer order parameter vanishes as well in the thermodynamic limit, being, for finite $N$,  considerably smaller  than the plaquette VBC one.  

In a recent investigation, based on a renormalized tensor network calculation,  plaquette-like order has been found in the thermodynamic limit.\cite{ren} We note that our GS energies   are lower than those obtained in Ref. \onlinecite{ren}. For example, for a system comprising $256$ spins at $J_2=0.5$ we find, with the simplest and economical EPS WF, based on $2\times2$ entangled plaquettes, $e^{EPS}(N=256, l=4, J_2=0.5)=-0.46299(3)$ being  $e^{RTN}(N=256, J_2=0.5)=-0.45062$ the value reported in Ref. \onlinecite{ren}.  

A VBC phase has also been obtained in Ref. \onlinecite{IM}. Such an ordered phase is a consequence  of to the relatively small bond dimension of the projected-entangled pair states adopted, in the mentioned work, to describe the the WF of the system, and is found to vanish when a larger bond dimension is used.\cite{peps}

In Ref. \onlinecite{edsq}, the authors cannot rule out the existence of a plaquette VBC phase, due to the difficulty of the extrapolation to the thermodynamic limit of the relevant order parameter. The study presented in Ref. \onlinecite{edsq} is based on the ED, in the subspace of short-range valence bond states, of the Hamiltonian  (\ref{ham}). This technique is applicable, due to the unfavorable scaling with the systems size of the computational cost, only to small lattices (i.e., comprising up to $N \sim 40$ spins). It is therefore not surprising that, in Ref. \onlinecite{edsq},  where results for finite lattices of maximum size $N=40$ are provided, an accurate and reliable extrapolation to the thermodynamic limit of the VBC order parameter  is not achievable.  Conversely, in this work, we use a general variational WF (i.e., the EPS one) which allows to study the GS properties of lattices whose size is considerable larger than that  presently accessible to  ED approaches. This is a crucial aspect,  as it  renders the extrapolation to the large $N$ limit of physical observables more accurate and reliable.

Having found that for  $0.5\lesssim J_2\lesssim0.6$  the GS of model (\ref{ham}) is a spin liquid, namely a paramagnet with no VBC order, we investigate the nature of such a spin liquid by computing the Renyi  entanglement entropy $S_2$. This quantity can be defined  in a form suitable to Monte Carlo estimation:\cite{ign}
\begin{figure}[b] 
{\includegraphics[scale=0.68]{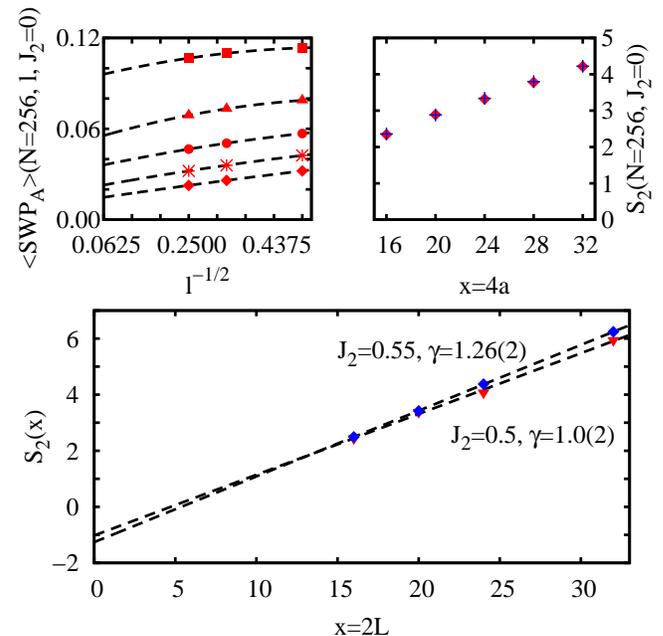}}
\caption{(color on line). Upper-left part: dependence of the Monte Carlo average of the swap operator on the plaquette size for square subregion $A$ comprising $a\times a$ spins. Different symbols refer to different $a$ values ($a=4$ to $8$ from top to bottom). Each dashed line is a quadratic (in $l^{-1/2})$ fit to numerical data with the same $a$. Upper-right part: extrapolated (to the $l\rightarrow N$ limit) values of the entanglement entropy as a function of the boundary length $x$ of subregion $A$ (diamonds). Also shown, for comparison, the QMC estimates (crosses)\cite{melko1}  Lower part: entanglement entropy for two values of $J_2$ at which the GS is a spin liquid. Each dashed line is a linear fit (to numerical data with the same $J_2$) performed to extract the topological entropy $\gamma$ (see text). Here $A$ is a $L/2\times L$ subregion of a $L\times L$ lattice with PBC. Hence, $A$ has no corners.} 
\label{fig:5}
\end{figure}
\begin{eqnarray}
e^{-S_2(A)}\!&=&\!\!\!\!\!\!\!\!\!\!\sum_{\mathbf{a}_1,\mathbf{b}_1,\mathbf{a}_2,\mathbf{b}_2}\!\!\!\!\!\!\!\!\!W^2(\mathbf{a}_1,\mathbf{b}_1)W^2(\mathbf{a}_2,\mathbf{b}_2)\frac{W(\mathbf{a}_2,\mathbf{b}_1)W(\mathbf{a}_1,\mathbf{b}_2)}{W(\mathbf{a}_1,\mathbf{b}_1)W(\mathbf{a}_2,\mathbf{b}_2)}\nonumber \\
&=& \langle SWP_A\rangle
\label{ent}
\end{eqnarray}
where $A$ is a portion of the system and $B$ its complement, $\mathbf{a}_i$  ($\mathbf{b}_i$), with $i=1,2$ labeling two copies of the system, denotes the spin configuration (e.g., along the $z$-axis) of the lattice sites belonging to subregion $A$ ($B$), $W$ is the weight of a global configuration, and $SWP_A=\frac{W(\mathbf{a}_2,\mathbf{b}_1)W(\mathbf{a}_1,\mathbf{b}_2)}{W(\mathbf{a}_1,\mathbf{b}_1)W(\mathbf{a}_2,\mathbf{b}_2)}$ is the  swap operator.\cite{melko1}

Figure \ref{fig:5} shows the Monte Carlo average of the swap operator, for a system of size $N=256$  at $J_2=0$, as a function of the plaquette sizes (upper-left part). Different symbols refer to square ($a\times a$) subregion $A$ comprising a different number of lattice sites. The $l \rightarrow N$ limit of $\langle SWP_A\rangle$, obtained by fitting  data with the same $a$ using a second order polynomial in $l^{-1/2}$ (dashed lines), defines, via Eq. (\ref{ent}), our estimates of the entanglement entropy (shown in the upper-right part of the same figure, as a function of the boundary length $x$ of region $A$).  Our results for $S_2$ are in excellent agreement with exact QMC ones (crosses).\cite{melko1} 
Using the same procedure we compute $S_2(x)$ at $J_2=0.5$ and $J_2=0.55$ (lower part of Fig. \ref{fig:5}), where the GS of the system is a spin liquid. In this case the subregion $A$ is a $L/2\times L$ portion of a $L\times L$ system, therefore, given that we are using PBC, it has no corners. A fit of our data assuming the scaling function $S_2(x)=dx-\gamma$ allows us to identify the topological entropy $\gamma$ which is finite (see figure) in both the cases studied here. We note that, at $J_2=0.5$, our estimated topological entropy, $\gamma=1.0(2)$, is consistent (taking into account the statistical uncertainty)  with that of a $\mathbb{Z}_2$ spin liquid,\cite{last} and in quantitative agreement with that computed in  recent studies based on a density matrix renormalization group calculation on long cylinders,\cite{dmrg} and on a tensor product approach using a cluster update imaginary time evolution method.\cite{peps}

\section{Conclusions}

We have studied the ground-state phase diagram of the spin-$1/2$ $J_1-J_2$ model on the square lattice by means of an entangled-plaquette  variational ansatz. Values of ground-state energy and relevant order parameters have been computed for $0\le J_2 \le 1$. In addition to magnetically ordered ground states occurring at ${J_2}\lesssim 0.5$ (N\'eel type),  and   $J_2 \gtrsim 0.6$ (collinear type), we find that the ground state, in the intermediate region of the phase diagram, is a topological spin liquid.  Further investigations, which are beyond the purpose of  this study,  are needed to understand the nature of the phase transition occurring between the N\'eel phase  and the spin liquid one as well as between the latter and the collinear phase. For example, the N\'eel to spin liquid appears to be a continuous phase transition. However, its full characterization is highly non trivial and a subject of great current  interest.\cite{last1} 
We want to emphasize that the numerical approach employed here allowed us to provide accurate quantitative predictions which, plausibly in the near future, might be  tested experimentally  by using ultracold atoms in an optical lattice, essentially as {\it quantum simulators} of the  $J_1-J_2$  Hamiltonian.\cite{last2} Finally, it has to be mentioned that the entangled-plaquettte variational wave function (as well the methodology employed in our study) can be applied, without substantial modification, to basically {\it any} lattice model. It gives a straightforward access to crucial observables such as magnetic and non magnetic order parameters or  the Renyi entanglement entropy, constituting, in our view, one of the most reliable variational choices to investigate the ground-state physics of frustrated (fermionic) Hamiltonians, intractable, due to the sign problem, with exact Quantum Monte Carlo techniques. 

\section*{Acknowledgements}
The author acknowledges discussions with J. I. Cirac, and thanks A. B. Kallin, and J. F. Yu for having provided the QMC results for the entanglement entropy, and the energy values obtained in Ref. \onlinecite{ren}. This work has been supported by the EU project QUEVADIS.


\begin{thebibliography}{20}
\bibitem{note} {Henceforth, we set the value of $J_1$ to unity, and take it as our energy unit.}
\end{thebibliography}
\end{document}